\documentclass[two column,linenumbers,superscriptaddress,groupedaddress amssymb, amsmath,nobibnotes,nofootinbib, aps, showpacs, prb]{revtex4}
\usepackage[dvips]{graphicx}
\usepackage{graphicx}

\begin{document}

\title{Broken time-reversal symmetry probed by muon spin relaxation in the caged type superconductor Lu$_5$Rh$_6$Sn$_{18}$}
\author{A.Bhattacharyya}
\email{amitava.bhattacharyya@stfc.ac.uk} 
\affiliation{ISIS Facility, Rutherford Appleton Laboratory, Chilton, Didcot Oxon, OX11 0QX, United Kingdom} 
\affiliation{Highly Correlated Matter Research Group, Physics Department, University of Johannesburg, PO Box 524, Auckland Park 2006, South Africa}
\author{D.T. Adroja} 
\email{devashibhai.adroja@stfc.ac.uk}
\affiliation{ISIS Facility, Rutherford Appleton Laboratory, Chilton, Didcot Oxon, OX11 0QX, United Kingdom} 
\affiliation{Highly Correlated Matter Research Group, Physics Department, University of Johannesburg, PO Box 524, Auckland Park 2006, South Africa}
\author{J. Quintanilla} 
\affiliation{ISIS Facility, Rutherford Appleton Laboratory, Chilton, Didcot Oxon, OX11 0QX, United Kingdom} 
\affiliation{SEPnet and Hubbard Theory Consortium, School of Physical Sciences, University of Kent, Canterbury CT2 7NH, United Kingdom}
\author{A. D. Hillier}
\affiliation{ISIS Facility, Rutherford Appleton Laboratory, Chilton, Didcot Oxon, OX11 0QX, United Kingdom} 
\author{N. Kase}
\affiliation{Department of Physics and Mathematics, Aoyama-Gakuin University, Fuchinobe 5-10-1, Sagamihara, Kanagawa 252-5258, Japan}
\author{A.M. Strydom} 
\affiliation{Highly Correlated Matter Research Group, Physics Department, University of Johannesburg, PO Box 524, Auckland Park 2006, South Africa}
\affiliation{Max Planck Institute for Chemical Physics of Solids, 01187 Dresden, Germany}
\author{J. Akimitsu} 
\affiliation{Department of Physics and Mathematics, Aoyama-Gakuin University, Fuchinobe 5-10-1, Sagamihara, Kanagawa 252-5258, Japan}

\date{\today}

\begin{abstract}
The superconducting state of the caged type  compound Lu$_5$Rh$_6$Sn$_{18}$ has been investigated by using magnetization, heat capacity, and muon-spin relaxation or rotation ($\mu$SR) measurements and the results interpreted on the basis of the group theoretical classifications of the possible pairing symmetries and a simple model of the resulting quasiparticle spectra. Our zero-field $\mu$SR measurements clearly reveal the spontaneous appearance of an internal magnetic field below the transition temperature, which indicates that the superconducting state in this material is characterized by the broken time-reversal symmetry. Further the analysis of  temperature dependence of  the magnetic penetration depth measured using  the transverse field $\mu$SR  measurements suggest an isotropic $s$-wave character for the superconducting gap. This is in agreement with the heat capacity behavior and we show that it can be interpreted in terms of a non unitary triplet state with point nodes and an open Fermi surface.

\end{abstract}
\maketitle

It is a major theoretical challenge in strongly correlated electron systems to understand the pairing mechanism in unconventional superconductors~\cite{jb,ms}. In conventional `$s-$wave' superconductors, only gauge symmetry is broken. If the pairing is not conventional then some other symmetries of the Hamiltonian may be broken below the superconducting transition. Symmetries which might be broken include lattice point and translation group operations and spin rotation symmetries, in addition to the global gauge symmetry that is responsible for the Meissner effect, flux quantization, and the Josephson effects. The nature of the broken symmetry in the pairing state is reflected in the symmetry properties of the order parameter. Superconductors whose crystal structure features a center of inversion, can be classified via the parity of Cooper pair state: the spin-singlet pair state ($S$ = 0) corresponds to an orbital pair wave function $\psi(k) \sim \psi(-k)$ with even parity [i.e., $\Delta(k) = \Delta(-k)$]; The spin-triplet state (total spin $S$ = 1) has a superconducting order parameter with odd parity[$\psi(k) \sim -\psi(-k)$]~\cite{jn}.  A few compounds have been reported to be spin-triplet superconductors, for example the 4$d$-electron system Sr$_2$RuO$_4$~\cite{gm,apm,ym1}, and the 5$f$-electron systems UPt$_3$~\cite{gml} and UNi$_2$Al$_3$~\cite{ki}.  

\par

Broken symmetry can modify the physics of a system and results in novel and uncommon behavior. Superconductivity is one of the finest illustrations of a symmetry breaking phenomenon.  A particularly interesting case is time$-$reversal symmetry (TRS) breaking. This is rare and has only been observed directly in a few unconventional superconductors, e.g., Sr$_2$RuO$_4$~\cite{gm,jx}, UPt$_3$~\cite{gml} and (U;Th)Be$_{13}$~\cite{rhh}, (Pr;La)(Os;Ru)$_4$Sb$_{12}$~\cite{ya}, PrPt$_4$Ge$_{12}$~\cite{am}, LaNiC$_2$~\cite{ad1}, LaNiGa$_2$~\cite{ad2} and Re$_6$Zr~\cite{rps}. A direct manifestation of broken TRS is the appearance of spontaneous weak magnetic fields, detected in these systems by zero field muon spin relaxation (ZF$-\mu$SR). ZF$-\mu$SR is useful to search for TRS breaking fields; the presence of such fields limits the possible superconducting states and the associated pairing symmetry. For example, TRS is a prerequisite for any state with a one-dimensional representation (singlet, triplet or admixed), and its breaking is associated with special kinds of states which have a degenerate representation. The presence of two or more degenerate superconducting phases naturally leads to a spatially inhomogeneous order parameter near the resulting domain walls; this creates spontaneous supercurrents and hence magnetic fields near those regions. Another possible origin of TRS-breaking fields is from intrinsic magnetic moments due to spin polarization (for spin- triplet pairing) and the relative angular momentum of the Cooper pairs~\cite{ms}.  Specifically one can prove, using group-theoretical arguments~\cite{ad2}, that non-unitary triplet pairing (thought to occur in noncentrosymmetric LaNiC$_{2}$~\cite{ad1} and centrosymmetric LaNiGa$_2$ ~\cite{ad2}) leads to a small bulk magnetization $M$. The latter acts as a sub-dominant order parameter of the superconducting instability i.e. it grows only linearly with decreasing temperature, $M\sim T_{\bf c}-T$~\cite{ad2}. Recently the size of this magnetization has been obtained within a non-unitary triplet pairing model of Sr$_2$RuO$_4$~\cite{km1}. 
\par
The possibility of singlet-triplet pairing in noncentrosymmetric superconductors makes them prime candidates to exhibit TRS breaking.  In spite of this, it is well established theoretically~\cite{jq1} and experimentally~\cite{eb1} that singlet-triplet mixing does not imply necessarily broken TRS. On the other hand broken TRS has been observed in Re$_6$Zr~\cite{rps} where we expect a strong singlet-triplet admixture. In contrast, for LaNiC$_2$ symmetry analysis implies that the superconducting instability is of the purely triplet type, with a spin-orbit coupling that is comparatively weak and with mixing of singlet and triplet pairing being forbidden by symmetry~\cite{jq}. 

\par
Caged type structures have received considerable attention due to their fascinating properties~\cite{zh}. Three cage compounds have been comprehensively studied over the past decade as ``rattling-good'' materials: Ge/Si clathrates, filled skutterudites (RT$_4$X$_{12}$), and $\beta-$pyrochlore oxides (AOs$_2$O$_6$)~\cite{zh}. Typically they possess three dimensional skeletons surrounding large atomic cages, inside of which reasonably small atoms are situated and can ``rattle'' with large atomic excursions due to the virtual size inconsistency, weak structural coupling, and strong electron$-$phonon (rattler) coupling, leading to a considerable anharmonicity for rattling vibration. For instance, rattling of the A atoms in the OsO$_6$ cages induce extremely strong-coupling superconductivity in AOs$_2$O$_6$~\cite{zh1}. A strong interplay between quadrupolar moment and superconductivity has been pointed out in RT$_4$X$_{12}$~\cite{kk} and RT$_{2}$X$_{20}$~\cite{to}. R$_5$Rh$_6$Sn$_{18}$ (R = Sc, Y, Lu), which can also be categorized as the cage compounds, exhibit superconductivity with the transition temperature $T_{\bf c}$ = 5 K (Sc), 3 K (Y), and 4 K (Lu)~\cite{jpr}. These compounds have a tetragonal structure with the space group $I4_1/acd$ and Z = 8, where R occupies two sites of different symmetry ~\cite{sm}. In this Rapid communications, we report on ZF$-\mu$SR and TF$-\mu$SR measurements for Lu$_{5}$Rh$_{6}$Sn$_{18}$. The results unambiguously reveal the spontaneous appearance of an internal magnetic field in the SC state, providing clear evidence for broken time reversal symmetry. 

\begin{figure}[t]
\vskip -0.0 cm
\centering
\includegraphics[width = 9.5 cm]{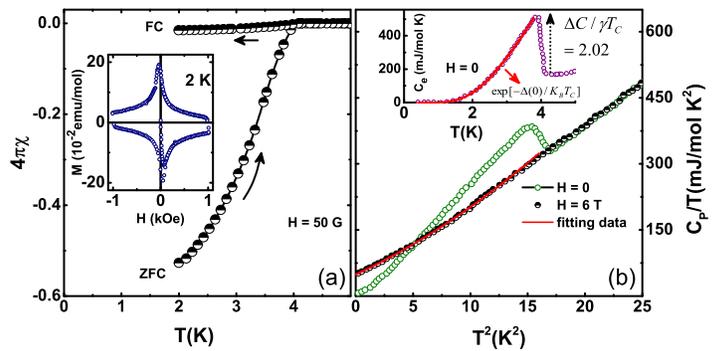}
\caption {(Color online) (a) The temperature dependence of the dc magnetic susceptibility of Lu$_{5}$Rh$_{6}$Sn$_{18}$. The inset in (a) isothermal field dependence of magnetization at 2.0 K. (b)  shows the $C_P/T$ vs. $T^2$ curve. The solid line shows the fit (see text).  The inset in (b) temperature dependent of electronic specific heat $C_e$ under zero field after subtracting the lattice contribution for Lu$_5$Rh$_6$Sn$_{18}$.}
\end{figure}

Single crystals of Lu$_5$Rh$_6$Sn$_{18}$ were grown by a conventional Sn-flux method in the ratio of Lu:Rh:Sn = 1:2:20. A detailed discussion on the crystal growth can be found in Ref.~\cite{jpr}.~Well defined Laue diffraction spots indicated the good quality of the single crystals with a typical size 3x3x3mm. Powder X-ray diffraction patterns were indexed as the Lu$_5$Rh$_6$Sn$_{18}$ phase with the space group $I4_1/acd$~\cite{jpr}. The magnetic measurements were performed using a Quantum-Design MPMS. Specific heat measurement were performed down to 500 mK by a relaxation method calorimeter (Quantum Design PPMS equipped with a $^3$He refrigerator). 

 \begin{figure}[b]
\vskip -0.0 cm
\centering
\includegraphics[width = 5.5 cm]{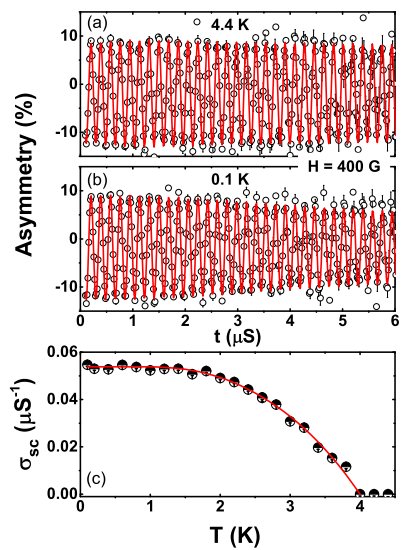}
\caption {(Color online) The transverse-field muon time spectra (one component) for Lu$_{5}$Rh$_{6}$Sn$_{18}$ collected (a) at $T$ = 4.4 K and (b) at $T$ = 0.1 K in a magnetic field $H$ = 400 G. (c) The temperature dependence of  $\sigma_{sc}(T)$. The line is a fit to the data using an isotropic model (Eq. 2).}
\end{figure} 

\par
Muon spin relaxation ($\mu$SR) experiment were carried out on the MUSR spectrometer at the ISIS pulsed muon source of the Rutherford Appleton Laboratory, U.K~\cite{sll}. The $\mu SR$ experiments were conducted in zero$-$field (ZF), longitudinal$-$field (LF), and transverse$-$field (TF) mode. High quality single crystal of Lu$_5$Rh$_6$Sn$_{18}$ was mounted on a sample plate made of 99.995\% sliver, which was placed in a dilution refrigerator with a temperature range of 100 mK to 4.5 K. Using an active compensation system the stray magnetic fields at the sample position were canceled to a level of 1 $\mu$T. TF$-\mu$SR experiments were performed in the superconducting mixed state in applied field 400 G, well above the $\mu_0$$H_{c1}$= 20 G of this material.~Data were collected in the field$-$cooled mode where the magnetic field was applied above the superconducting transition and the sample was then cooled down to base temperature. Muon spin relaxation is a dynamic method to resolve the type of the pairing symmetry in superconductors~\cite{js}. The mixed or vortex state in case of type-II superconductors gives rise a spatial distribution of local magnetic fields; which demonstrates itself in the $\mu$SR signal through a relaxation of the muon polarization.
\par

Magnetization measurement indicate that Lu$_5$Rh$_6$Sn$_{18}$ is a bulk superconductor with a superconducting transition temperature $T_{\bf c}$ = 4.0$\pm$(0.1) K  as shown in Fig. 1 (a). Below $T_{\bf c}$ the low$-$field $\chi(T)$ shows a robust diamagnetic signal.  The shielding volume fraction is $\sim$53\% at 2 K. Inset of Fig. 1 (a) shows the magnetization $M(H)$ curve at 2 K, which is typical for type-II superconductivity. Resistivity [$\rho(T)$, not shown here] exhibits a very unusual temperature variation~\cite{nk1}. $\rho(T)$ is nearly independent of $T$ down to about 120 K, and shows an increase on further cooling~\cite{nk1}. Fig. 1 (b) shows the $C_P(T)$  at $H$ = 0 and 6 T. At 4.0 K a sharp anomaly is observed indicating the superconducting transition which matches well with $\chi(T)$ data. Since the normal-state specific heat was found to be invariant under external magnetic fields, the normal-state electronic specific heat coefficient $\gamma$ and the lattice specific heat coefficient $\beta$ were deduced from the data in a field of 6 T by a least-square fit of the $C_P/T$ data to $C_P/T = \gamma +\beta T^2 + \delta T^4$. The least squares analysis of the 6 T data provides a Sommerfeld constant $\gamma$ = 48.10$\pm$(0.5) mJ/(mol-K$^2$), $\delta$ = 0.32$\pm$(0.03) mJ/(mol K$^6$) and the Debye temperature $\Theta_D$ = 157$\pm$(2) K. We obtained the specific heat jump $\Delta C_P(T_C)$ = 397$\pm$(3) mJ/(mol K) and $T_{\bf c}$ = 4.0$\pm$(0.2) K, which yields  $\Delta C$/$\gamma T_C$ = 2.06$\pm$(0.03). From the exponential dependence of $C_e$ as shown in the inset of Fig. 1 (b), we obtained 2$\Delta$(0)/$k_B$$T_C$ to be 4.26$\pm$(0.04). Because this value is relatively larger than that of the theoretical BCS limit of weak-coupling superconductor (3.54), this compound can be categorized as a strong-coupling superconductor~\cite{nk2}. 

\begin{figure}[t]
\vskip -0cm
\centering
\includegraphics[width = 8.0 cm]{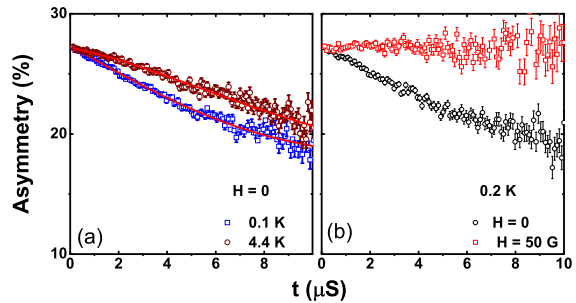}
\caption {(Color online) (a) Zero-field $\mu$SR time spectra for Lu$_5$Rh$_6$Sn$_{18}$ collected at 0.1 K (square) and 4.4 K (circle) are shown together with lines that are least squares fits to the data using Eq. (3). These spectra collected below and above $T_{\bf c}$ are
representative of the data collected over a range of  $T$. (b) A LF$-\mu$SR time spectrum taken in an applied field of  5 mT at 0.2 K is also shown.}
\end{figure}

\par
Fig. 2 (a) and (b) show the TF$-\mu$SR precession signals above and below $T_{\bf c}$ with an applied field of 400 G (well abobe $H_{c1}$). Below $T_{\bf c}$ the signal decays with time due to inhomogeneous field distribution of the flux-line lattice. The TF$-\mu$SR asymmetry spectra were fitted using an oscillatory decaying Gaussian function,

\begin{equation}
\begin{split}
G_{z1}(t) = A_1cos(2\pi \nu_1 t+\phi_1)exp\left({\frac{-\sigma^2t^2}{2}}\right)\\ + A_2cos(2\pi \nu_2 t+\phi_2)
\end{split}
\end{equation}

where $\nu_1$ and $\nu_2$ are the frequencies of the muon precession signal and background signal, respectively, $\phi_i$ ($i$ = 1,2) are the initial phase offset.  The first term gives the total sample relaxation rate $\sigma$; there are contributions from both the vortex lattice ($\sigma_{sc}$) and nuclear dipole moments ($\sigma_{nm}$, which is assumed to be constant over the entire temperature range) below $T_{\bf c}$ [ where $\sigma$ = $\sqrt{(\sigma_{sc}^2+\sigma_{nm}^2)}$]. The contribution from the vortex lattice, $\sigma_{sc}$, was determined by quadratically subtracting the background nuclear dipolar relaxation rate obtained from spectra measured above $T_{\bf c}$. As $\sigma_{sc}$ is directly related to the magnetic penetration depth, the superconducting gap can be modeled by,

\begin{equation}
\frac{\sigma_{sc}(T)}{\sigma_{sc}(0)}=\frac{\lambda^{-2}(T)}{\lambda^{-2}(0)}=1+2\int_{\Delta(T)}^{\infty}\left (\frac{\delta f}{\delta E}\right) \frac{EdEd\phi}{\sqrt{E^2-\Delta(T)^2}}
\end{equation}

where $f = [1+exp(-E/K_B T)]^{-1}$ is the Fermi function~\cite{mt}. The temperature dependence  of the gap is approximated by the expression $\delta(T/T_C)$ =tanh$\{1.82[1.018(T_C/T-1)]^{0.51}\}$~\cite{ac}.

\begin{figure}[t]
\vskip -0 cm
\centering
\includegraphics[width = 8 cm]{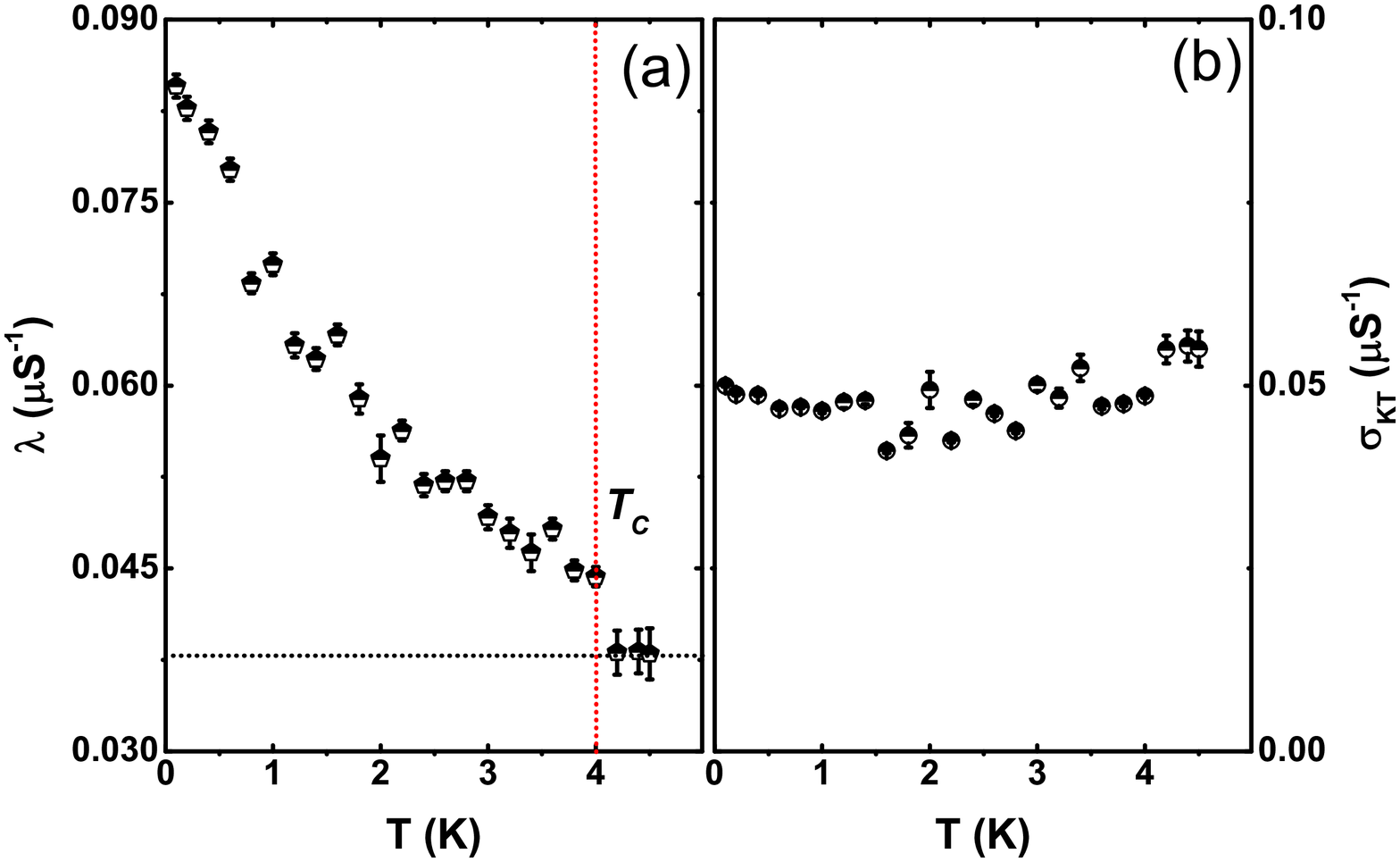}
\caption {(Color online) (a) The temperature dependence of the electronic relaxation rate measured in zero magnetic field of Lu$_5$Rh$_6$Sn$_{18}$ with $T_{\bf c}$ = 4.0 K is shown. The lines are guides to the eye. The extra relaxation below $T_{\bf c}$ indicates additional internal magnetic fields and, consequently, suggests the superconducting state has broken time reversal symmetry. (b) The Kubo-Toyabe  depolarization rate $\sigma_{KT}$, versus temperature in zero field shows no temperature dependence.}
\end{figure}

\par
Fig. 2 (c) shows the $T$ dependence of the $\sigma_{sc}$ which can be directly related to the superfluid density. From this, the nature of the superconducting gap can be determined. The data can be well modeled by a single isotropic gap of 0.75$\pm$ 0.06 meV. This gives a gap of 2$\Delta$/$k_B$$T_{\bf c}$ = 4.4$\pm$0.02, which is higher than the 3.53 expected for BCS superconductors. This is a further indication of the strong electron-phonon coupling in the superconducting state. Lu$_5$Rh$_6$Sn$_{18}$ is a type II superconductor, assuming that roughly all the normal state carriers ($n_e$) contribute to the superconductivity (i.e., $n_s\approx n_e$), we have estimated the values of effective mass of the quasiparticles $m^*\approx 1.32 m_e$ and  superconducting electron density $\approx$ 2.6 $\times$10$^{28}$ m$^{-3}$ respectively. More details on these calculations can be found in Ref. \cite{adsd,vkasd,dtasd}.

\par
The time evolution of the ZF$-\mu$SR is shown in Fig. 3 (a) for $T$ = 100 mK and 4.4 K. In these relaxation experiments, any muons stopped on the silver sample holder give a time independent background. No signature of precession is visible, ruling out the presence of a sufficiently large internal magnetic field as seen in magnetically ordered compounds. The only possibility is that the muon$-$spin relaxation is due to static, randomly oriented local fields associated with the nuclear moments at the muon site. The ZF$-\mu$SR data are well described by the damped Gaussian Kubo-Toyabe (KT) function,

\begin{equation}
G_{z2}(t) =A_1 G_{KT}(t)e^{-\lambda t}+A_{bg}
\end{equation}

where $G_{KT}(t) =\left[\frac{1}{3}+\frac{2}{3}(1-\sigma_{KT}^2t^2)e^{{\frac{-\sigma_{KT}^2t^2}{2}}}\right]$;  $\lambda$ is the electronic relaxation rate, $A_1$ is the initial asymmetry, $A_{bg}$ is the background.  The parameters $\sigma_{KT}$ [Fig. 4 (b)], $A_1$, and $A_{bg}$ are found to be temperature independent. It is remarkable that $\lambda$ shows a significant increase [Fig. 4 (a)] with an onset temperature of 4.0$\pm$0.1 K, indicating the appearance of a spontaneous internal field correlated with the superconductivity.  This observation provides unambiguous evidence that TRS is broken in the SC state of Lu$_5$Rh$_6$Sn$_{18}$. Such a change in $\lambda$ has only been observed in superconducting Sr$_2$RuO$_4$~\cite{gm}, LaNiC$_2$~\cite{ad1} and SrPtAs~\cite{pkb}.  This increase in $\lambda$ can be explained in terms of a signature of a coherent internal field with a very low frequency as discussed by Luke {\it et. al.}~\cite{gm} for Sr$_2$RuO$_4$. This suggest that the field distribution is Lorentzian in nature similar to Sr$_2$RuO$_4$. Considering similar temperature dependence of $\lambda$ in Sr$_2$RuO$_4$, LaNiC$_2$, SrPtAs and  Lu$_5$Rh$_6$Sn$_{18}$, we attribute this behavior of $\lambda$ to the TRS breaking below $T_{\bf c}$ in Lu$_5$Rh$_6$Sn$_{18}$.  A longitudinal magnetic field of just 50 G [Fig. 3 (b)] removes any relaxation due to the spontaneous fields and is sufficient to fully decouple the muons from this relaxation channel. This in turn shows that the associated magnetic fields are in fact static or quasistatic on the time scale of the muon precession. These observations further support the broken TRS in the superconducting state of Lu$_5$Rh$_6$Sn$_{18}$. The increase in the exponential relaxation below $T_{\bf c}$ is, 0.045 $\mu$S$^{-1}$, which corresponds to a characteristic field strength $\lambda/\gamma_\mu$=  0.5 G. This is about the same as we observed in the B phase of UPt$_3$ and Sr$_2$RuO$_4$~\cite{gml}. No theoretical estimates of the characteristic field strength in Lu$_5$Rh$_6$Sn$_{18}$ are yet available; however, we expect them to be comparable to those in Sr$_2$RuO$_4$ and UPt$_3$ as the fields should arise from a similar mechanism.

\begin{figure}
\vskip -0.0 cm
\centering
\includegraphics[width=9cm]{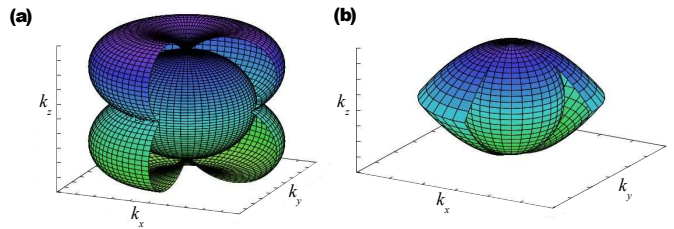}
\caption{\label{gapfig}Nodal structure of the allowed singlet (a) and triplet (b) pairing states. See Supplementary Online Material for details.}
\label{figapshort}
\end{figure}

\par
Our main observation, namely the breaking of TRS on entering the superconducting state, has important implications for the symmetry of pairing and for the quasi-particle spectrum. In short, a standard symmetry analysis~\cite{jfa,Sig} carried out under the assumption of strong spin orbit coupling, yields two possible pairing states, one with $d+id$ character (singlet) and another one non-unitary (triplet). As shown in Fig.~\ref{figapshort}, both states are nodal: the singlet has a line node and two point nodes, and the triplet has two point nodes. At temperatures $T \ll T_c$, the thermodynamics of the singlet state would be dominated by the line node, yielding for example $C \sim T^2$ for the specific heat. Similarly, the triplet state would be dominated by the point nodes, which happen to be shallow (a result protected by symmetry) and therefore also lead to $C \sim T^2$~\cite{Mazi}. However, because of the location of the nodes in the triplet case, fully-gapped behavior may be recovered depending on the topology of the Fermi surface. Moreover some limiting cases of the triplet state correspond to regular, i.e. linear point nodes ($C \sim T^3$) as well as to a more exotic state with a nodal surface (gapless superconductivity, $C \sim T$). Finally, a fully-gapped spectrum will result if the Fermi surface is open at the locations of the point nodes. The allowed pairing states and their quasiparticle spectra are discussed in detail in the Supplementary Online Material. We note that the theoretical analysis presented there is valid for any superconductor with $D_{4h}$ point group symmetry, strong spin-orbit coupling and broken time-reversal symmetry and may therefore be applied for example to Sr$_2$RuO$_4$~\cite{Veen}, as well as Lu$_5$Rh$_6$Sn$_{18}$. 

\par
In conclusion, we have used both ZF$-\mu$SR and TF$-\mu$SR to investigate the superconductivity of the cage type tetragonal system Lu$_5$Rh$_6$Sn$_{18}$. The ZF$-\mu$SR measurements show a spontaneous field appearing at the superconducting transition temperature. The presence of spontaneous internal magnetic fields in our measurements suggests that a time-reversal symmetry breaking mixed symmetry pairing state does occur below $T_{\bf c}$. TF$-\mu$SR measurements yield a magnetic penetration depth that is exponentially flat at low temperatures, and so our data can be fit to a single-gap BCS model. Symmetry analysis suggests either a singlet $d+id$ state with a line node or, alternatively, nonunitary triplet pairing with point nodes, which may be linear or shallow and can become fully gapped depending on the Fermi surface topology. 

\par
We would like to thank Dr M. Smidman for help in $\mu$SR data analysis. A.B would like to acknowledge FRC of UJ, NRF of South Africa and ISIS-STFC for funding support. DTA and ADH would like to thank CMPC-STFC, grant number CMPC-09108, for financial support.  AMS thanks the SA-NRF (Grant 78832) and UJ Research Committee for financial support. JQ gratefully acknowledges financial support from STFC and from HEFCE through the South-East Physics network (SEPnet). JQ thanks Paul Strange, Phil Whittlesea for useful discussions.

\pagebreak

\part*{\bf\Large\center Broken time-reversal symmetry probed by muon spin relaxation in the caged type superconductor Lu$_5$Rh$_6$Sn$_{18}$: \\ SUPPLEMENTAL MATERIAL}

\begin{center}
{A.Bhattacharyya$^{1,2}$}
{D.T. Adroja$^{1,2}$} 
{J. Quintanilla$^{1,3}$} 
{A. D. Hillier$^{1}$}
{N. Kase$^{4}$}
{A.M. Strydom$^{2,5}$} 
{J. Akimitsu$^{4}$} 

\vspace{0.5cm}

$^1${\it ISIS Facility, Rutherford Appleton Laboratory, Chilton, Didcot Oxon, OX11 0QX, UK} 

$^2${\it Highly Correlated Matter Research Group, Physics Department, University of Johannesburg, PO Box 524, Auckland Park 2006, South Africa}

$^3${\it SEPnet and Hubbard Theory Consortium, School of Physical Sciences, University of Kent,\\ Canterbury CT2 7NH, UK}

$^4${\it Department of Physics and Mathematics, Aoyama-Gakuin University, Fuchinobe 5-10-1, Sagamihara, \\Kanagawa 252-5258, Japan}

$^5${\it Max Planck Institute for Chemical Physics of Solids, 01187 Dresden, Germany}

\end{center}

\twocolumngrid

\begin{figure}
\centering
\includegraphics[width=1.0\columnwidth]{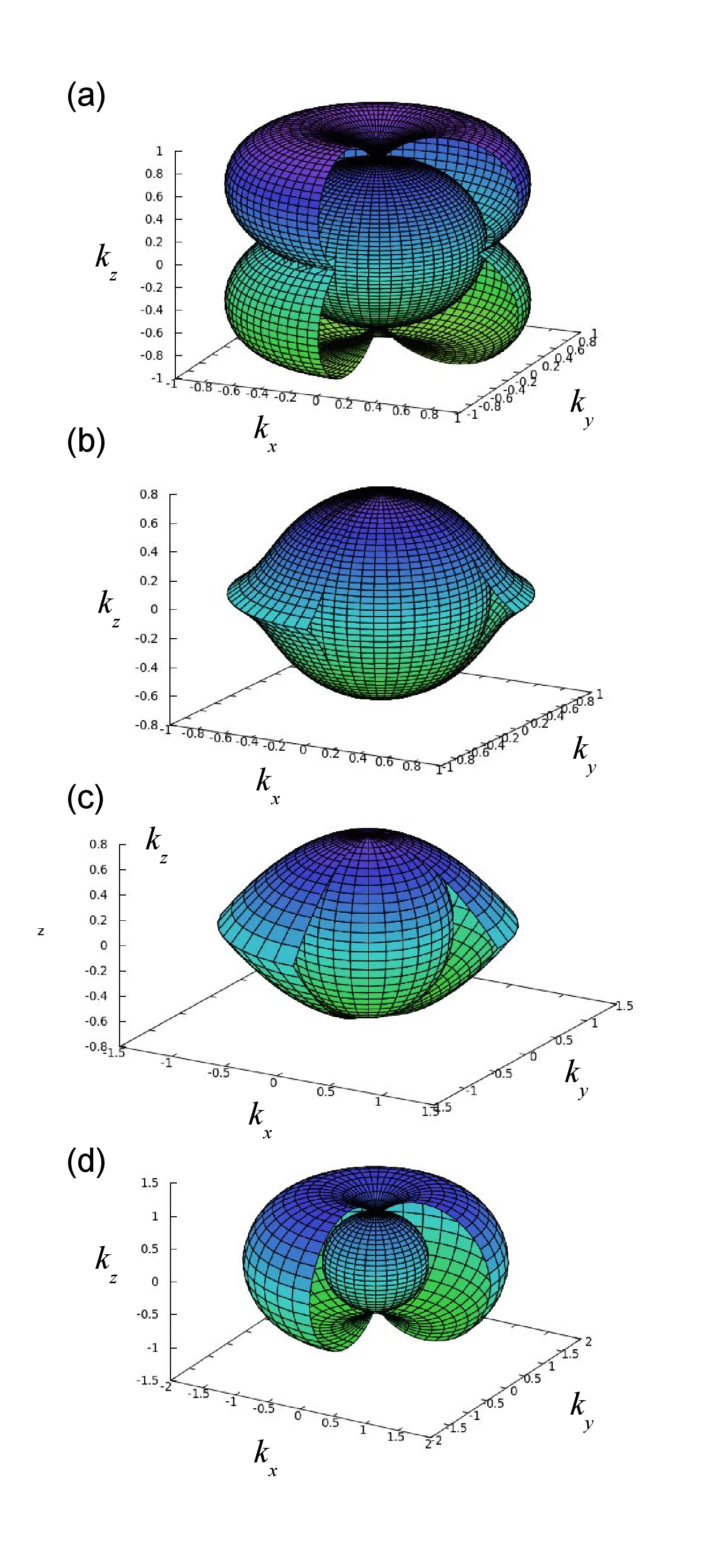}
\caption{\label{gapfig}Four possible forms of the gap in the quasi-particle energy spectrum, plotted on a spherical Fermi surface. (a) Singlet gap function [Eq.~(\ref{singap})] showing a line node on the equator and two linear point nodes at the ``North'' (N) and ``South'' (S) poles. (b-d) The same but for the triplet gap function [Eq.~(\ref{trigap})] with $B \ll A$, $B \sim A$ and $B \gg A$, respectively. In all cases we assumed the functions $X,Y,Z$ in Eqs.~(\ref{singap},\ref{trigap}) to take their simplest forms: $k_x,~k_y$ and $k_z$, respectively.}
\label{figap}
\end{figure}

Here we expound in detail our group-theoretical arguments about the symmetry of the superconducting order parameter. We also present the associated nodal structure of the quasiparticle spectrum. The analysis presented here applies to any superconductor with $D_{4h}$ point group symmetry, broken time-reversal symmetry and strong spin-orbit coupling. Besides Lu$_5$Rh$_6$Sn$_{18}$, the ruthenate superconductor Sr$_2$RuO$_4$ has been recently argued to fall within this category \cite{Veenstra2014}. We note, however, that our present analysis does not allow for singlet-triplet mixing as put forward in reference ~\cite{Veenstra2014}.

Barring an independent magnetic transition whose critical temperature is fine-tuned to coincide with the superconducting critical temperature, the sudden increase in the muon spin relaxation at $T_c$ suggests that the superconducting state breaks time-reversal symmetry. Assuming that the superconductivity is static and does not break the translational symmetry of the lattice, it can be characterised by a momentum-dependent pairing potential $\Delta_{\alpha,\beta}({\bf k})$. Here $\alpha,\beta=\uparrow$ or $\downarrow$ are the spin indices of the two electrons in a Cooper pair and $\hbar{\bf k}$ is the momentum of one of the electrons with respect to the centre of mass of the pair. Standard symmetry analysis \cite{Annett1990,Sigrist1991} yields $\Delta_{\alpha,\beta}({\bf k}) = \sum_{n=1}^D \Delta_n \Gamma^n_{\alpha,\beta}({\bf k})$ just below $T_c$, where the functions ${\Gamma^n}_{n=1,2,\ldots,D}$ form a basis set of one of the irreducible representations of the point group of the crystal, of dimension $D$. The form of the coefficients $\Delta_n$ is obtained by minimizing generic free energies of the appropriate symmetry. The superconducting state breaks time-reversal symmetry if $D>1$ and two or more coefficients have different complex phases  \cite{Annett1990}. The quasiparticle spectrum is then given by the diagonalisation of the Bogoliubov-de Gennes Hamiltonian
\begin{equation}
H_{\rm BdG} = \left( 
\begin{matrix}
	\epsilon({\bf k})-\mu	& 0 			& \Delta_{\uparrow\uparrow}({\bf k}) 				& \Delta_{\uparrow\downarrow}({\bf k})			\\
	0 				& \epsilon({\bf k})-\mu	& \Delta_{\downarrow\uparrow}({\bf k}) 				& \Delta_{\downarrow\downarrow}({\bf k})			\\
	\Delta^*_{\uparrow\uparrow}({\bf k}) 				& \Delta^*_{\downarrow\uparrow}({\bf k}) 			& -\epsilon({\bf k})+\mu 	& 0			\\
	\Delta^*_{\uparrow\downarrow}({\bf k})	 				& \Delta^*_{\downarrow\downarrow}({\bf k})	 			& 0				& -\epsilon({\bf k})+\mu\\
\end{matrix}
\right),
\end{equation}
where $\epsilon({\bf k})$ is the single-electron dispersion relation, measured from the chemical potential. Note that in noncentrosymmetric systems (not considered here) the off-diagonal elements in the single-electron dispersion relation would be essential. 

For the space group I$4_1$/acd the relevant point group is tetragonal $D_{4h}$~\cite{ITC} which has been thoroughly examined in the context of cuprate superconductivity~\cite{Annett1990,Sigrist1991}. Let us first consider the case of singlet pairing. Quite generally, $\hat{\Delta}({\bf k})=\Delta_0({\bf k})i\hat{\sigma}_{y}$ (where $\hat{\sigma}_y$, is the second Pauli matrices). Following Ref.~\cite{Annett1990}, for the point group of interest, there are 7 possible instabilities, only one of which (corresponding to the $^1E_g(c)$ irrep) breaks time-reversal symmetry. The gap function in this case has the form \cite{Annett1990} 
\begin{equation}
	\Delta_0({\bf k})=(X+iY)Z
	\label{singap}
\end{equation} 
where $X,Y,Z$ are three real functions of ${\bf k}$ that transform as $k_x$, $k_y$ and $k_z$, respectively, under the point group symmetry operations. Fig.~\ref{gapfig}~(a) depicts the size, at the Fermi surface, of the corresponding gap in the quasiparticle energy spectrum, $\propto \Delta_0({\bf k})$, obtained by assuming an isotropic single-electron dispersion relation, $\epsilon({\bf k}) = \hbar^2|{\bf k}|^2/2m^*$ (yielding a spherical Fermi surface) and  taking the simplest forms for the functions $X,Y,Z$, namely $k_x$, $k_y$ and $k_z$, respectively.  The gap is given as follows
\begin{equation}
\Delta({\bf k}) \propto |k_z|\sqrt{k_x^2+k_y^2} 
\end{equation}
and so the low energy excitations would be dominated by a line node at the equator ($k_z=0$), leading to a specific heat proportional to $T^2$ at low temperatures $T\ll T_c$ (the only exception being if the Fermi surface does not traverse the $k_z=0$ plane, which is unlikely). In addition, there are point nodes at the ``North'' (N) and ``South'' (S) poles of the Fermi surface. 

In the case of triplet pairing, i.e.~$\hat{\Delta}({\bf k})=i\left[{\bf d}({\bf k}).\hat{\boldsymbol{\sigma}}\right]\hat{\sigma}_{y}$, the double group combining the point group of the crystal with spin rotations has to be considered. Still following \cite{Annett1990}, for $D_{4h}$ we find once more that only one of 7 possible instabilities (corresponding to the $E_u(c)$ irrep of the double group) breaks time-reversal symmetry. The ${\bf d}$-vector in this case is given by \cite{Annett1990}
\begin{equation}
{\bf d}({\bf k})=(AZ,iAZ,B(X+iY))
\label{trigap} 
\end{equation}
where the real coefficients $A$ and $B$ depend on details of the band structure and effective electron-electron interactions. This gap function corresponds to non-unitary triplet pairing as ${\bf d}^* \times {\bf d} \neq 0$.

The gap for this triplet case, with the same simplifying assumptions used above, is depicted in Fig.~\ref{gapfig} (b-d). Its formula is 
\begin{equation}
	\Delta({\bf k}) \propto 
		\left| 
			A|k_z| - \sqrt{A^2 k_z^2 + B^2\left(k_x^2+k_y^2\right)}
		\right| 
\end{equation}
Interestingly, the spectrum also features two point nodes at $X=Y=0$, but in this case the point nodes are ``shallow'' using the terminology of \cite{Mazidian2013}. Indeed this instability is analogous to the $E_{2u}$  instability proposed for UPt$_3$ which also features a shallow node \cite{Joynt2002}. These nodes become ordinary linear nodes when $B \gg A$, while they expand to cover the whole Fermi surface in the opposite limit, $B \ll A$ (gapless superconductivity). Thus we expect the power-law exponent $n$ characterising the low-temperature behaviour of the specific heat, $C \sim T^n$, to be $1$, $2$, and $3$ for the cases represented in panels (b), (c) and (d), respectively.

Non-unitary triplet pairing leads to the breaking of the two-fold degeneracy between the spin-up and spin-down parts of the quasi-particle spectrum. In this case we expect the superconducting instability to be accompanied by a bulk magnetisation that grows linearly with $T_c-T$ for $T \lesssim T_c$ and which acts as a sub-dominant order parameter \cite{Hillier2012,Miyake2014}. The linear increase of the muon spin relaxation rate $\lambda$ that we observe experimentally (see Fig. 4 of the main text) also increases linearly below $T_c$, which suggests that $\lambda$ is simply proportional to this (very small \cite{Miyake2014}) bulk magnetisation.

We emphasise that the power laws mentioned above are only expected to be realised in the limit of very low temperatures ($T \ll T_c$). Moreover, if the topology of the Fermi surface departs significantly from a sphere the spectrum may become fully-gapped i.e. $C \sim e^{-\Delta/T}$ for $T \ll T_c$. Specifically, the triplet pairing potential may lead to a fully-gapped spectrum if the Fermi surface is open at the top and the bottom, so that it never cuts the N and S poles. This would lead to temperature dependences of specific heat, superfluid density, etc.  like those of an $s-$wave superconductor in spite of the broken time-reversal symmetry. In contrast the singlet order parameter would lead to line nodes with $C \sim T^2$ at low temperatures always. Finally, we point out that a broader range of possibilities emerge if spin-orbit coupling happens to be weak enough to be neglected \cite{Annett1990}. In that case two or more non TRS-breaking instabilities may merge to give new TRS-breaking ones \cite{Quintanilla2010}.

\end{document}